\documentclass{article}

\usepackage{amsmath}
\usepackage{amssymb}
\usepackage{authblk}
\usepackage{graphicx}
\usepackage{float}
\usepackage{epstopdf}
\usepackage{geometry}

\title{Heun-type solutions for Schwarzschild metric with electromagnetic fields}
\author[1]{T. Birkandan}
\author[2]{M. Horta\c{c}su}
\affil[1]{Istanbul Technical University - Department of Physics, Istanbul, Turkey}
\affil[2]{Mimar Sinan Fine Arts University - Department of Physics, Istanbul, Turkey}

\begin{document}
\maketitle

\begin{abstract}
We find confluent Heun solutions to the radial equations of two Halilsoy-Badawi metrics. For the first metric, we studied the radial part of the massless Dirac equation and for the second case, we studied the radial part of the massless Klein-Gordon equation.
\end{abstract}

\section{Introduction}
Heun functions \cite{Heun, Ronveaux} seem to be still a novelty
among theoretical physicists although they were introduced nearly
130  years ago. After the centennial conference which took place
in 1989 and the papers presented in this conference were published
in a book \cite{Ronveaux}, there was an explosion of papers in
this field \cite{Hortacsu}. Many equations whose exact solutions
were not known turned out to have solutions in this set. After
referring to people \cite{Blaudin,Leaver, Suzuki} who tried to
show whether the exact solutions of the celebrated ``Teukolsky
Master Equation" \cite{Teukolsky} can be written down
in terms of the confluent forms of the Heun equation, Batic and
Schmidt \cite{Batic} showed that the ``Teukolsky master Equation" (and similar equations) could be transformed in any
relevant type-D metric into a Heun form. Although the Heun
equation and its confluent forms are  much better known today in
the theoretical physics community and included in some
mathematical packages, we still find some authors who do not
identify the equations they find properly.

Here we give two examples of metrics which yield confluent Heun
solutions for the equations describing  a test particle
whose wave equation is written in the background metric of these
metrics in the linear approximation, ignoring the backreaction
and nonlinear terms in the Einstein-Maxwell equations.
In the first example the massless Dirac equation and in the second,
the massless Klein-Gordon equation are studied.

\section{Dirac equation}
In a very interesting paper \cite{Badawi} Al-Badawi and Owaidat study the Dirac equation in the background of the spherically symmetric solution of the Einstein- Maxwell equations, analogous to the Schwarzschild metric, in the presence of spherically symmetric static electromagnetic field.

For the metric they use a solution found by one of these authors
with Halilsoy \cite{Halil}. Actually this metric
was previously discovered by  Ray and  Wei\cite{Ray}. Metrics, when
a Schwarzschild metric is in an homogeneous electromagnetic field
were given by Halilsoy \cite{Halil1, Halil2}, as stated in
\cite{Badawi2}, and were included in the book by Griffiths and
Podolsky \cite{podolsky}. This solution is a superposition of the
Schwarzschild \cite{Sch} solution with an external, stationary
electromagnetic Bertotti-Robinson solution \cite{ Bert, Rob}.

The metric is given as
\begin{eqnarray}
ds^{2}&=&\frac{r^{2}-2Mr}{r^{2}f(r)}\left[ dt-Mq(1+a^{2})\cos \theta d\phi \right] ^{2} \nonumber \\
&&-\frac{r^{2}f(r)}{r^{2}-2Mr}dr^{2}-r^{2}f(r)\left( d\theta
^{2}+\sin ^{2}\theta d\phi ^{2}\right),
\end{eqnarray}
\begin{eqnarray}
r^2f(r) =\frac{1}{2}r\left( r-2M\right) \left[ p\left(
1+a^{2}\right) +a^{2}-1\right] +2Mar+M^{2}\left[ p\left(
1+a^{2}\right) -2a\right].
\end{eqnarray}
This is a type-D metric. $M$ is a constant parameter
which mimics the role of the source mass in Newtonian
approximations for the geodesics of test particles in the
Schwarzschild metric when one considers large values of the
luminosity radius $r$; $p$ is the twisting parameter of the
external electromagnetic field and $a$ is the interpolation
parameter between two metrics used. This metric is in
the class named as Plebanki and Demianski solutions\cite{PD}.

When the parameter $a$ is zero one gets a metric which
can be transformed into the Bertotti- Robinson solution by setting
the twisting parameter to one. As stated in \cite{podolsky}, in
general this space time is a direct product of two two dimensional
spaces of constant curvature, namely the 2-sphere and two-
dimensional anti-de Sitter space-times. When $a$ is between zero
and one, it contains a family of expanding
Schwarzschild-Reissner-Nordstrom-de Sitter metrics. Here, as in
\cite{Ray,Halil} we take the cosmological constant equal to zero.

The authors consider only a spinor test particle, in the linear
approximation, ignoring the backreaction and nonlinear terms in
the Einstein-Maxwell equations. They use the Newman-Penrose
formalism \cite{Ted} to separate the Dirac equation into radial
and angular ($\theta$) parts after assuming a periodic solution
for the variables $t$ and $\phi$ which are along Killing
directions. They can solve the angular equation in terms of the
associated Legendre functions. For the radial equation, they use
the WKB approximation and write their solutions in terms of
exponentials.

We just considered the case in which the coupled spinor
test particle has zero mass. When the test particle is massive, we
could not obtain the exact solution in terms of known functions.
For the massless case, the radial equations are of the form
\cite{Badawi}:
\begin{eqnarray}
\frac{dT_{1}}{dr}+i\frac{kR^{2}}{H^{2}R^{2}}T_{1}=\frac{\lambda}{RH}T_{2}, \\
\frac{dT_{2}}{dr}-i\frac{kR^{2}}{H^{2}R^{2}}T_{2}=\frac{\lambda}{RH}T_{1}.
\end{eqnarray}
Here $\lambda$ is the eigenvalue, $R^2 = r^2 f(r)$
and
\begin{equation}
H^{2}=\frac{r^{2}-2Mr}{r^{2}f\left( r\right) }.
\end{equation}
The exact solutions for the radial equations above turn out to be
confluent Heun functions  multiplied by exponentials and powers of
the finite regular singular points of the radial equation. This is
no surprise, since as Philipp and Perlick pointed out
\cite{Philipp}, Leaver\cite{Leaver1,Leaver} and
Fiziev \cite{Fiziev,Fiziev1,Fiziev3} showed that  the solution of a
particle in the Schwarzschild background results in confluent Heun
type solutions.
In this respect, it looks like the solutions of
the Eguchi-Hanson instanton \cite{EH}  trivially extended to five
dimensions \cite{Tolga}.  In this case, too,  the angular equation
has a solution which is in the family of hypergeometric functions,
whereas the radial equation's solution is again a confluent Heun
function \cite{Heun,Ronveaux}.

In the paper of Al-Badawi and Owaidat \cite{Badawi}, we could not find the explicit
expressions for the frequency $\omega$, the essential parameter in
the WKB approximation, aside from an integral, to check the
behavior of the incoming and outgoing waves as well as the
expressions for the transition and reflection coefficients are not
given. The authors end their paper by studying the plots of the
effective potentials for different values of the parameters in the
effective potential.

One can show that the radial equation can indeed be solved in
terms of confluent Heun functions. This solution has regular
singularities at $r=0$ and $r=2M$ where the horizon is, and an
irregular singularity at infinity. We first study the solution for
$0<r<2M$, assuming that the given metric is also valid here. We
are aware of the fact that the metric used is only valid outside
the event horizon due to the properties of the Schwarzschild
metric. We do this  calculation just to compare our solution with
that given in \cite{Badawi} and see if it has the
correct behavior around the regular singular points of the
differential equation.

We did not give the second order equation obtained from the system
of first order equations given above for the sake of brevity of
the paper, since that equation is quite long. We, however, give
a similar equation when we study the region for $r$ greater than
$2M$, the domain of $r$  in which we are really interested in this paper.

We only take the first solution which is analytic around the
singularity at $r=0$. It reads as:
\begin{eqnarray}
T_1(r) &=& {{\rm e}^{i/2 \left(  \left( p+1 \right) {a}^{2}+p-1 \right) rk}} {r}^{i/2k \left( {a}^{2}p-2\,a+p \right) M
} \left( 2\,M-r \right) ^{1/2+i/2k \left( {a}^{2}p+2\,a+p \right) M} \times \nonumber \\
&&{H_C} \bigg( 2\,iM \left( \left( p+1 \right) {a}^{2}+p-1 \right) k,-1/2+ik \left( {a}^{2}p-2\,a+p \right) M, \nonumber \\
&& 1/2+ik \left( {a}^{2}p+2\,a+p \right) M, \left( 4\,Mak+i \right) M \left(  \left( p+1 \right) {a}^{2}+p-1 \right) k, \\ \nonumber
&& 1/2\,{k}^{2}{M}^{2} \left( {a}^{2}+1 \right) ^{2}{p}^{2} \\
&&-1/2\,M \left( {a}^{2}+1 \right)  \left( -2\,M{a}^{2}k+4\,Mak+2\,kM
+i \right) kp-2\,{k}^{2}{a}^{3}{M}^{2} \nonumber \\
&&-1/2\,Mk \left( -4\,kM+i
 \right) {a}^{2}+M \left( 2\,kM+i \right) ka+i/2kM-{\lambda}^{2}+3/8,
{\frac {r}{2M}} \bigg), \nonumber
\end{eqnarray}
where $H_C$ denotes the confluent Heun function. We note that
although this solution is analytic around $r=0$, it is not
analytic around $r=2M$. This is expected, since our solution is only a  local solution which is analytic only in the neighborhood of one singularity \cite {Ronveaux2}. Here the parameters are defined in \cite{Badawi}.

These equations also have a second solution:
\begin{eqnarray}
T_{1_2}(r) &=& {{\rm e}^{i/2rk \left(  \left( p+1 \right) {a}^{2}+p-1 \right) }}{r}^{
1/2-i/2k \left( {a}^{2}p-2\,a+p \right) M} \left( 2\,M-r \right) ^{1/2
+i/2 \left( {a}^{2}p+2\,a+p \right) kM} \times \\ \nonumber
&&{H_C} \bigg( 2\,iM \left(  \left( p+1 \right) {a}^{2}+p-1 \right) k,1/2-ik \left( {a}^{2}p-2\,a+p \right) M, \\ \nonumber
&& 1/2+i \left( {a}^{2}p+2\,a+p \right) kM, M \left(
 \left( p+1 \right) {a}^{2}+p-1 \right)  \left( 4\,kMa+i \right) k, \\ \nonumber
&& 1/2\,{M}^{2}{k}^{2} \left( {a}^{2}+1 \right) ^{2}{p}^{2} \\ \nonumber
&& -1/2\,M \left( {a
}^{2}+1 \right)  \left( -2\,M{a}^{2}k+4\,kMa+2\,kM+i \right) kp-2\,{k}
^{2}{a}^{3}{M}^{2} \\ \nonumber
&& -1/2\,M \left( -4\,kM+i \right) k{a}^{2}+M \left( 2
\,kM+i \right) ka+i/2Mk-{\lambda}^{2}+3/8,{\frac {r}{2M}}
 \bigg). \nonumber
\end{eqnarray}
In the following, we do not consider it, since it has a square
root irregularity at $r=0 $, our point of expansion.

The standard form of the confluent Heun equation is given as \cite{Fiz1,Fiz2}
\begin{equation}
    {\frac{{d^{2}H_C}}{{dz^{2}}}}+\left( \alpha +{\frac{{\gamma+1}}{{z-1}}}+{\frac{{\beta+1}}{{z}}}%
    \right) {\frac{{dH_C}}{{dz}}}+\left({\frac{{\mu}}{{z}}}+ {\frac{{\nu}}{{z-1}}}\right)H_C =0, \label{fizheun}
\end{equation}
with solution $H_C(\alpha, \beta, \gamma, \delta,\eta,z)$, and the parameters have the relations
\begin{equation}
    \delta = \mu+\nu-\alpha \bigg( \frac{{ \beta+\gamma+2}}{{2}} \bigg),
\end{equation}
\begin{equation}
    \eta = \frac{{ \alpha(\beta+1) }}{{2}} - \mu - \frac {{ \beta+\gamma+ \beta \gamma}}{{2}}.
\end{equation}
Since we are interested in the region $r>2M$, outside the event horizon, the solution we gave above does not
suit our purposes if we want to investigate the behavior of the wave for $r>2M$. To find a solution
to suit our purpose, we have to transform to the variable  $u= r-2M$. This will give us one solution which is analytic around $r=2M$.
This solution may not be analytic around $r=0$. Since we are not interested in the region $0<r<2M$, this will not cause any problems.

From the two first order differential equations given above, equations [3] and [4], we derive a second order equation for $F_1$. This equation reads
\begin{equation}
A {\frac {{\rm d}^{2}}{{\rm d}{u}^{ 2}}}{\it T_1} \left( u \right)
+ B {\frac {\rm d}{{\rm d}u}}{\it T_1} \left( u \right) + \left(
C+D+E)\right) {\it T_1} \left( u \right) = 0,
\end{equation}
where
\begin{equation}
A= \left( u+2\,M \right) ^{2}{u}^{2},
\end{equation}
\begin{equation}
 B=\left( M+u \right) \left( u+2\,M
 \right) u,
\end{equation}
\begin{eqnarray}\nonumber
C&=&[  \left(  \left( p/2+1/2 \right) {a}^{2}+p/2-1/2
 \right)  \left( u+2\,M \right) ^{2}- \left(  \left( p+1 \right) {a}^{
2}-2\,a+p-1 \right) M \left( u+2\,M \right) \\ &+& {M}^{2} \left(
{a}^{2}p-2 \,a+p \right)] ^{2}{k}^{2},
\end{eqnarray}
\begin{eqnarray}\nonumber
D &=&[\left(  \left( i/2+i/2p
 \right) {a}^{2}+i/2p-i/2 \right)  \left( u+2\,M \right) ^{3}-3/2\,iM
 \left(  \left( p+1 \right) {a}^{2}+p-1 \right)  \left( u+2\,M
 \right)^{2} \\ &+& i \left( a-1 \right) {M}^{2} \left( a+1 \right)  \left(
u+2\,M \right) +i{M}^{3} \left( {a}^{2}p-2\,a+p \right) ]k,
\end{eqnarray}
\begin{equation}
E={\lambda}^{2} \left( u+2\,M \right) u.
\end{equation}
Here we again take the solution which is analytic around $u=0$
($r=2M$), namely
\begin{eqnarray}
T_1(u) &=&{{\rm e}^{-i/2uk \left(  \left( p+1 \right) {a}^{2}+p-1 \right) }} \left( u+2\,M \right)^{i/2k \left( {a}^{2}p-2\,a+p \right) M}{u}^{-i/2 \left( {a}^{2}p+2\,a
+p \right) kM} \times \nonumber \\
&&{H_C} \bigg( 2\,iM \left(  \left( p+1 \right) {a}^{2}+p-1
 \right) k,-1/2-i \left( {a}^{2}p+2\,a+p \right) kM, \nonumber \\
&& -1/2+ik \left( {a}
^{2}p-2\,a+p \right) M,-M \left(  \left( p+1 \right) {a}^{2}+p-1
 \right)  \left( 4\,kMa+i \right) k, \\
&& 1/2\,{M}^{2}{k}^{2} \left( {a}^{2}
+1 \right) ^{2}{p}^{2}+1/2\,M \left( {a}^{2}+1 \right)  \left( 2\,M{a}
^{2}k+4\,kMa-2\,Mk+i \right) kp \nonumber \\
&& +2\,{k}^{2}{a}^{3}{M}^{2}+1/2\,M
 \left( 4\,Mk+i \right) k{a}^{2}+Mk \left( -2\,Mk+i \right) a-i/2Mk-{
\lambda}^{2}+3/8,-{\frac {u}{2M}} \bigg) . \nonumber
\end{eqnarray}
The second solution is given below:
\begin{eqnarray}
T_{1_2}(u) &=&{{\rm e}^{-i/2uk \left(  \left( p+1 \right) {a}^{2}+p-1 \right) }}
 \left( u+2\,M \right) ^{i/2k \left( {a}^{2}p-2\,a+p \right) M}{u}^{1/
2+i/2 \left( {a}^{2}p+2\,a+p \right) kM} \times \\ \nonumber
&& {H_C} \bigg( 2\,iM
 \left(  \left( p+1 \right) {a}^{2}+p-1 \right) k,1/2+i \left( {a}^{2}
p+2\,a+p \right) kM, \\ \nonumber
&&-1/2+ik \left( {a}^{2}p-2\,a+p \right) M,-M
 \left(  \left( p+1 \right) {a}^{2}+p-1 \right)  \left( 4\,kMa+i
 \right) k, \\ \nonumber
&& 1/2\,{M}^{2}{k}^{2} \left( {a}^{2}+1 \right) ^{2}{p}^{2}+1/
2\,M \left( {a}^{2}+1 \right)  \left( 2\,M{a}^{2}k+4\,kMa-2\,kM+i
 \right) kp \\ \nonumber
&& +2\,{k}^{2}{a}^{3}{M}^{2}+1/2\,M \left( 4\,kM+i \right) k{a
}^{2}+kM \left( -2\,kM+i \right) a-i/2Mk-{\lambda}^{2}+3/8,-{\frac {u}{2M}} \bigg). \nonumber
\end{eqnarray}
We discard it since it has a second root non analyticity at $u=0$,
our point of expansion.

Using $p=10$, $k=0.2$, $a=0.1$, $\lambda=0.7$ and $M=5$, we give the plots of the first solution for  $0<r<2M$ and $u>0$ in Figure 1 and Figure 2,
respectively.
\begin{figure}[H]
\centering
\includegraphics[scale=0.3]{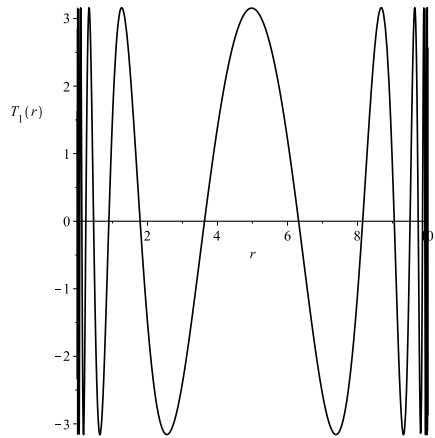} \caption{Solution between $0<r<2M$}
\label{fig:fig1}
\end{figure}
\begin{figure}[H]
\centering
\includegraphics[scale=0.3]{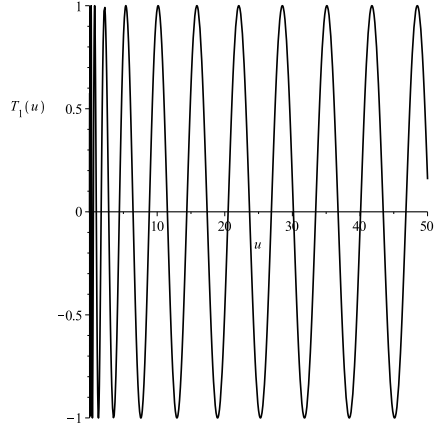} \caption{Solution for $u>0$}
\label{fig:fig2}
\end{figure}
Polynomial solutions can be given for the confluent Heun equation under some conditions \cite{Fiz2,ciftci}. The identity $\mu+\nu=-N\alpha$,
$N$ being the degree of the polynomial solution, should be satisfied along with a vanishing determinant. However, this identity is not useful
in this case as $\mu+\nu=0$.

\vspace{5pt}
\noindent \textbf{Solution around infinity}

Solution of the confluent Heun equation around the irregular singularity at infinity can be given by the Thom{\'e} solution as \cite{figu}
\begin{eqnarray}
\lim_{z\rightarrow \infty}U(z)\sim e^{\pm i\omega z}z^{\mp i\eta-(B_{2}/2)}.
\end{eqnarray}
Note that this is just the first term in a nonconverging series, which must be considered as just an asymptotic one.

The confluent Heun equation is written in the form
\begin{eqnarray}
z(z-1)\frac{d^{2}U}{dz^{2}}+(B_{1}+B_{2}z)\frac{dU}{dz}+\left[B_{3}-2\eta\omega(z-1)+\omega^{2}z(z-1)\right]U=0,
\end{eqnarray}
for $\omega\neq0$ and all other parameters are constants. This
form is called the generalized spheroidal wave equation and
finding the correspondence with the general form given by the
equation (\ref{fizheun}) needs some algebra as the solutions are
not in the same form ({\it i.e.} we need to define
$V(z)=e^{i\omega z}U(z)$ first and then proceed with the
solution). Studying the solution for our case, we find the first term in this asymptotic series as
\begin{equation}
\lim_{u\rightarrow \infty}T_1(u) \sim e^{(2i[(p+1)a^2+p-1])Mku},
\end{equation}
as the solution around infinity. There is a second  solution which
behaves as $u^{4ikMa-1}$, i.e. it  vanishes as $u$ goes to
infinity.

As to the physical interpretation of our solutions, we
can get information only by plotting our solutions, since the
general behavior of Heun functions is not generally known
explicitly. We see the approach to singularity at $r=2M$ when we
plot our function for the range $0<r< 2M$. When we expand around
one of the regular singular points, we expect such a behaviour
around the second singular point.

To our surprise, the regular solution resembles a plane wave for
$u>0$ (Figure \ref{fig:fig2}) with almost constant frequency and
constant amplitude. This was a surprise, since this is the
behaviour only in the asymptotic region for the quasi-classical
solution in reference [9].  The irregular solution, besides being
non analytic at $u=0$, goes to  zero, oscillating with vanishing
amplitude.  This behaviour is reflected in the behaviour of the
second solution (asymptotic solution) which behaves as a reciprocal
power of $u$.
\section{Klein-Gordon equation}
In another recent paper by Al-Badawi, the Dirac equation is studied in a Schwarzschild
black hole immersed in an electromagnetic universe with charge coupling \cite{Badawi2}. Here the electromagnetic radiation is not attributed to the parameter M. This solution again interpolates the Schwarzschild \cite{Sch} and Bertotti-Robinson \cite{Bert,Rob} solutions \cite{Badawi2}. The metric is
\begin{equation}
ds^{2}=\frac{\Delta }{r^{2}}dt^{2}-\frac{r^{2}}{\Delta }dr^{2}-r^{2}\left(
d\theta ^{2}+\sin ^{2}\theta d\phi ^{2}\right) ,
\end{equation}
where $\Delta =r^{2}-2Mr+M^{2}\left( 1-a^{2}\right)$
\cite{halilbadawi,Ovgun} . Here, $M$ is the parameter
used in the Schwarzschild solution and $a$ $(0<a\leq 1)$
is the external parameter. This metric has two horizons, the outer
horizon  at $ r_{1} = M(1+a)$ and the inner horizon  at $ r_{2} =
M(1-a) $. The external electromagnetic field shrinks at the outer
horizon and expands at the inner horizon \cite{halilbadawi}. Note
that the $a=0$ case can be transformed into the Bertotti-Robinson
solution \cite{Ovgun}.

We study the massless Klein-Gordon equation in the background of
this  metric, since we can not identify the solution for
the massive case.
\begin{equation}
\frac{1}{\sqrt{-g}}\partial _{\mu }\left( \sqrt{-g}g^{\mu \nu }\partial_{\nu }\Phi \right)=0,
\end{equation}
in this background, namely,
\begin{eqnarray}
\Phi \sin^{2}\theta
  {r}^{4}{\omega}^{2}+ {\Delta} ^{2} \sin^{2}\theta{\frac {
\partial ^{2} \Phi}{\partial {r}^{2}}}  +{\Delta}  \left( {\frac {\partial \Phi}{\partial r}}
   \right)  \sin^{2}\theta{\frac {\rm d{\Delta}}{{\rm d}r}} -{\Delta}{{\it n}}^{
2}\Phi+{\Delta} \left( {\frac {\partial \Phi}{\partial
\theta}}  \right) \sin \theta
 \cos \theta  +{\Delta}
 \sin^{2} \theta {\frac {\partial ^{2} \Phi}
{\partial {\theta}^{2}}}=0.
\end{eqnarray}
This equation can be separated into the radial and angular parts
with the Ansatz
\begin{equation}
\Phi=e^{-i\omega t}e^{in\phi}F(r)S(\theta).
\end{equation}
After defining the separation constant $\lambda$, the
radial and angular parts are obtained as
\begin{eqnarray}
{\frac {{\rm d}^{2}{F\left( r \right)}}{{\rm d}{r}^{2}}}  -{
\frac {{F} \left( r \right) { \lambda}_{{}}}{{\Delta}
 }}+{\frac {{F} \left( r \right) {r}^{4}{\omega}^{2
}+{\Delta} \left( {\frac {\rm d{F(r)}}{{\rm d}r}}  \right) {\frac {\rm d{\Delta}}{{\rm d}r}}
 }{{\Delta} ^{2}}}=0, \\
{\frac {{\rm d}^{2} {S} \left( \theta
 \right)}{{\rm d}{\theta}^{2}}} +{S} \left( \theta \right) {\lambda}_{{}}+{\frac {
 \left( {\frac {\rm d {S} \left( \theta
  \right)}{{\rm d}\theta}}  \right) \sin \theta \cos \theta -{S} \left( \theta \right) {{n}}^{2}}{
\sin^{2} \theta }}=0.
\end{eqnarray}
The angular part is in the form of the associated Legendre equation and the radial part can be solved in terms of confluent Heun functions. We change our parameter $r$ to
$u=r-r_1$, $r_1$ being the  outer event horizon in order to study the behavior outside the event horizons. We note that the event horizon is located at $r_1=M(1+a)$ and the inner horizon is located at $r_2=M(1-a)$. The radial solution is
\begin{eqnarray}
F(u)&=&{{\rm e}^{-i\omega u}}{u}^{{\frac {i{{r_1}}^{2} \omega}{{r_1}-{r_2}}}} \left( u
+{r_1}-{r_2} \right) ^{{\frac {i{{r_2}}^{2} \omega}{{r_1}-{r_2}}}} \times   \\
&&{H_C} \bigg( 2\,i \omega \left( {r_1}-{r_2}
 \right) ,{\frac {2\,i{{r_1}}^{2} \omega}{{r_1}-{r_2}}},{\frac {2
\,i{{r_2}}^{2} \omega}{{r_1}-{r_2}}}, \left( -2\,{{r_1}}^{2}+2
\,{{r_2}}^{2} \right) {\omega}^{2}, \nonumber \\
&&{\frac {2\,{{r_1}}^{4}{\omega}^{2}-4\,{
{r_1}}^{3}{\omega}^{2}{r_2}-{\lambda}_{}{{r_1}}^{2}+2\,{r_1}
\,{r_2}\,{\lambda}_{}-{\lambda}_{}{{r_2}}^{2}}{ \left( {r_1}-{
r_2} \right) ^{2}}},-{\frac {u}{{r_1}-{r_2}}}
 \bigg), \nonumber
\end{eqnarray}
and the second solution, namely
\begin{eqnarray}
F_2(u)&=&{{\rm e}^{-i\omega u}} {u}^{{\frac {-i{{r_1}}^{2} \omega}{{r_1}-{r_2}}}} \left(
u+{r_1}-{r_2} \right) ^{{\frac {i{{r_2}}^{2} \omega}{{r_1}-{r_2}}}} \times \\ \nonumber
&& {H_C} \bigg( 2\,i\omega \left( {r_1}-{r_2}
 \right) ,{\frac {-2\,i{{r_1}}^{2} \omega}{{r_1}-{r_2}}},{\frac {2
\,i{{r_2}}^{2}\omega}{{r_1}-{r_2}}}, \left(-2\,{{r_1}}^{2}+2
\,{{r_2}}^{2} \right) {\omega}^{2}, \\ \nonumber &&{\frac
{2\,{{r_1}}^{4}{\omega}^{2}-4\,{
{r_1}}^{3}{r_2}\,{\omega}^{2}-{{r_1}}^{2}{\lambda}_{{}}+2\,{r_1}\,{r_2}\,{\lambda}_{{}}-{{r_2}}^{2}{\lambda}_{{}}}{
\left( {r_1}-{r_2} \right) ^{2}}},-{\frac {u}{{r_1}-{r_2}}} \bigg)
\nonumber
\end{eqnarray}
These solutions  may be interpreted as two waves with different
phases, but both moving in the same direction asymptotically,
since  for large values of $u$, $\ln(u)$ is much smaller than $u$.

We just wanted to state that this test particle in this metric,
too, has a Heun family solution. We will study other properties
of this solution  in further papers.
\section{Conclusion}
Here we studied two different metrics given by \cite{Halil} and \cite{halilbadawi}.
In the first case we studied the Dirac equation given
in \cite{Badawi} and found that the radial solution can be expressed in terms of confluent Heun functions.
We found the same structure in the second metric case \cite{Badawi2, halilbadawi} for the Klein-Gordon equation.

\section{ Acknowledgement} We thank the anonymous referee for correcting our ``careless" use of the physical and mathematical terminology. M.H. thanks  Prof. Ibrahim Semiz for
providing important literature and Prof.  Nadir Ghazanfari for technical assistance. He also thanks the Science Academy, Turkey for support. This work is supported by TUBITAK, the Scientific and Technological Council of Turkey.

\end{document}